\documentclass[12pt,preprint]{aastex}

\slugcomment{January 6, 2008.}

\shorttitle{Mid-IR PL Relations}
\shortauthors{Freedman {\it et al.}}

\begin{document}


\title{The Cepheid Period-Luminosity Relation at Mid-Infrared
Wavelengths: I. First-Epoch LMC Data}


\author{\bf Wendy L. Freedman, Barry F. Madore, Jane Rigby, S. E.  
Persson}
\affil{Observatories of the  Carnegie Institution of Washington \\ 813
Santa
Barbara St., Pasadena, CA ~~91101}
\email{wendy@ociw.edu, barry@ociw.edu, jrigby@ociw.edu,
persson@ociw.edu}
\author{\bf \& Laura Sturch}
\affil{Harvey Mudd College \\ Claremont CA ~~91711}
\email{laura.sturch@hmc.edu}



\begin{abstract}
We present the first mid-infrared Period-Luminosity (PL) relations for
Large Magellanic Cloud (LMC) Cepheids. Single-epoch observations of 70
Cepheids were extracted from Spitzer IRAC observations at 3.6, 4.5,
5.8 and 8.0~$\mu$m, serendipitously obtained during the SAGE
(Surveying the Agents of a Galaxy's Evolution) imaging survey of the
LMC. All four mid-infrared PL relations have nearly identical slopes
over the period range 6 - 88 days, with a small scatter of only
$\pm$0.16~mag independent of period for all four of these wavelengths.
We emphasize that differential reddening is not contributing
significantly to the observed scatter, given the nearly two orders of
magnitude reduced sensitivity of the mid-IR to extinction compared to
the optical.  Future observations, filling in the light curves for
these Cepheids, should noticably reduce the residual scatter.  These
attributes alone suggest that mid-infrared PL relations will provide a
practical means of significantly improving the accuracy of Cepheid
distances to nearby galaxies.

\end{abstract}


\vfill\eject
\section{INTRODUCTION}

Distance measurements to the Large Magellanic Cloud (LMC) have
historically played a critical role in the calibration of the
extragalactic distance scale (Feast \& Walker 1987; Madore \& Freedman
1991; Udalski {\it et al.} 1999; Freedman {\it et al.} 2001; Sandage
{\it et al.} 2006). The LMC Cepheid Period-Luminosity relations have
been adopted as the fiducial calibration sample for many recent
extragalactic distance measurements and for the determination of the
Hubble constant ({\it e.g.,} Freedman {\it et al.} 2001, Sandage {\it
et al.} 2006, Riess {\it et al.} 2005).

A number of different methods have been used to measure the distance
to the LMC. As tabulated in Gibson (2000) and Freedman {\it et al.}
(2001), most of the values for the LMC distance modulus based on these
different methods fall between 18.1 to 18.7 mag, corresponding to a
full range of 42 to 55 kpc, and likely reflecting the effects of
systematic errors in various methods. More recent values have tended
to cluster around a modulus of 18.5 mag (see Alves 2004; and also an
interesting commentary by Schaefer 2007)
\footnote[1]{For a regularly updated compilation of published
distances to the Large Magellanic Cloud see the web site maintained by
Ian Steer and Barry Madore through NED at
http://nedwww.ipac.caltech.edu/level5/NED0D/LMC\_ref.html.}.

In an effort to reduce the systematic errors in the LMC distance,
Persson {\it et al.} (2004) obtained near-infrared J, H, and K$_s$
photometric measurements of 92 LMC Cepheids. These stars were chosen
to be distributed across the LMC, with periods ranging from 3 to 100
days. They were also selected to be relatively isolated so that
crowding effects would be minimized. The sample also does not
contain overtone pulsators.  On average, 22 phase points were obtained
at each wavelength for each star. The dispersions in the infrared PL,
PLC, and extinction-free period-Wesenheit relations were found to be
very small, amounting to less than $\pm$0.14~mag, or 7\% in distance.

There are a number of advantages to obtaining observations of Cepheids
at long wavelengths (see, for example, the reviews by Madore \&
Freedman (1991, 1998) and references therein): (1) The sensitivity of
surface brightness to temperature is a steeply declining function of
wavelength. (2) The interstellar extinction curve decreases as a
function of increasing wavelength (being almost linear with
1/$\lambda$ at optical and near-infrared wavelengths). (3) At the
temperatures typical of Cepheids, metallicity effects predominate in
the UV, blue and visual parts of the spectrum where most of the line
transitions occur, with declining effects at longer
wavelengths (Bono {\it et al.} 1999). The overall insensitivity of
infrared magnitudes of Cepheids to each of these effects results in
decreased amplitudes for individual Cepheids, as well as a
decreased scatter in the apparent PL relations (first noted by
Wisniewski \& Johnson 1968, and by McGonegal {\it et al.} 1982,
respectively).

Given the pivotal role of the LMC distance in calibrating the
extragalactic distance scale, developing techniques to further reduce
systematic errors in that path remains a critical goal. The
mid-infrared capability of Spitzer offers a new opportunity to provide
a distance modulus to the LMC completely freed from the effects of
reddening and with decreased sensitivity to metallicity.  As we
discuss below, beyond about 4.5~$\mu$m, the resulting extinction
is at least a factor of 25 times smaller than the corresponding  
extinction
in the B band.

We report here on single-epoch, mid-infrared data for a subset of the
Persson {\it et al.} (2004) LMC sample, which were observed with
Spitzer at 3.6, 4.5, 5.8 and 8.0~$\mu$m. Data for 70 of the 92 Persson
et al. Cepheids were serendipitously obtained in the course of the
SAGE project (Meixner {\it et al.} 2006a), a large-area study of the
interstellar medium and current star formation in the LMC. Since
Cepheids are relatively young stars, large numbers of Cepheids were
also observed in these fields, and fortunately, as we show here, they
were observed with sufficient signal-to-noise to be suitable for our
purposes also. The single-epoch, mid-IR Cepheid PL relations resulting
from these observations have a total scatter of only $\pm$0.16~mag
(8\% in distance) for a single Cepheid with less than 1\%
(statistical) uncertainty in the mean for this sample. We discuss the
prospects for reducing the scatter even further.

\section{IRAC Mid-Infrared Period-Luminosity Relations}

SAGE, the Spitzer survey of the Large Magellanic Cloud is an acronym
for {\it Surveying the Agents of a Galaxy's Evolution} (Meixner {\it
et al.} 2006a). The survey covers a 7 x 7 degree region of the LMC
using both the IRAC and MIPS detectors, operating at 3.6, 4.5, 5.8,
8.0~$\mu$m, and 24, 70, 160~$\mu$m, respectively.  The MIPS
far-infrared data were obtained to probe the diffuse dust emission in
the LMC, while the IRAC data were obtained primarily to study the
stellar content; in the context of this study, we have analyzed only
the IRAC data. Two epochs of data have been obtained; however, at the
time of writing SAGE had released catalogs based on the first epoch
data only.


Catalogs of resolved sources have been prepared by the SAGE project
and made available through the Infrared Science Archive (IRSA). We
used the interface, GATOR, to retrieve the individual data
files manually by object name based on a position-based cone search
using the NED name resolver. Our in-going list of objects consisted of
the 92 Cepheids with extensive JHK$_s$ photometry published by Persson
et al. (2004). Only those Cepheids found in the catalog with periods
between 0.8 $<$ log(P[days]) $<$ 1.8 and having photometry at all four
wavelengths were retained. This sample of 70 Cepheids is listed in
Table 1, with the logarithm of the period in days (from Persson {\it
et al.}), plus magnitudes and errors in each mid-infrared
bandpass. The photometry is from Meixner {\it et al.} (2006a) and was
obtained using PSF fitting with a modified version of DAOPHOT, using
an iterative technique to measure the local background. Magnitudes are
on the Vega system.


The four mid-IR period-luminosity relations, based on single-epoch
observations ranging from 3.6 to 8.0~$\mu$m, are shown in Figures
1 and 2. SAGE mapped with half-array offsets, so that each epoch contains
two 12-second integrations and thus the total IRAC integrations were
24 seconds for each epoch. For 139 calibration stars, the SAGE team
compared their catalog magnitudes to those predicted from 2MASS fluxes
(Meixner {\it et al.}  2006a). The systematic offsets are $\sim$0.01
mag, and the standard deviation is $\sim$0.05 mag in all four IRAC
channels (Meixner {\it et al.}  2006b).  Since the fainter calibrators
have fluxes typical of our brighter Cepheids, $\pm$0.05~mag is a
reasonable expectation for the typical 1-$\sigma$ uncertainty for our
Cepheid photometry.  This expectation is consistent with the quoted
SAGE photometric uncertainties for our Cepheid sample, listed in Table
1.

As can be seen from Figures 1 and 2, the mid-infrared Period-Luminosity
relations: (1) all show small scatter, which is similar in magnitude
from filter to filter, (2) all four relations have very similar
slopes, and (3) as can be seen from Figure 3, the residuals around the
fits are highly correlated (with approximately unit slope) from one PL
relation to the next.  We note that the first two of these statements
do not apply to the visual and blue PL relations. But before turning
to address why this is the case, we first quantify the above
statements.

\subsection{Period-Luminosity Fits}

Weighted, least-squares, linear fits to each of the four mid-IR data
sets are given in \S 3. The slopes, zero points, respective errors
and $rms$ scatter are shown for each bandpass. The slopes at each
wavelength all have values of around -3.4, with a slight trend of the
slopes increasing to longer wavelengths. The scatter around each of
the fits is constant at $\pm$0.16~mag; and, from filter to filter the
scatter of individual data points is highly correlated (see Figure 3
and discussion below).

According to Gieren, Moffett \& Barnes (1999) the best current
estimate of the slope of the Period-Radius relation, using radial
velocity studies of Magellanic Cloud and Milky Way Cepheids is 
log(R) = 0.680 ~log(P) + C . If we convert log(R) to an area and then
express it as a magnitude, the slope derived from the Period-Radius
relation (0.680 $\pm$ 0.017) $\times$ -5 $=$ 3.40 $\pm$ 0.085 is
statistically indistinguishible from the longest-wavelength mid-IR PL
slopes of 3.44$\pm$0.03 and 3.42$\pm$0.03 at 8.0 and 5.8~$\mu$m,
respectively. This agreement suggests that the mid-IR PL relation is
in fact the Period-Area relation at fixed surface brightness. All of
these mid-IR wavelengths have small sensitivities to temperature,
which may also explain the similarity of slopes, as well as the small
magnitude of the scatter.

In Figure 4 we show the run of PL slopes as a function of increasing
wavelength, from the optical to the mid-infrared. The optical (BVRI)
slopes are from Madore \& Freedman (1998); near-infrared (JHK) slopes
are from Persson et al. (2004); and mid-IR slopes are from the present
paper. Following the dramatic change seen at optical wavelengths, the
slope appears to be asymptotically approaching a value of about -3.45,
which is within the currently published uncertainties for pure radius
variations (i.e. 3.40 $\pm$ 0.085). The latter is shown by the
horizontal lines crossing the bottom of Figure 4.

\subsection{Correlations in the Scatter}

The correlated nature of the small residual scatter in the mid-IR can
be explained by three contributors. For an intrinsically uniform
distribution of data points (as is the case for a Cepheid light curve)
the variance is formally equal to the range (i.e., the full amplitude)
divided by 12. If the typical amplitude of a Cepheid in the mid-IR is
comparable to the amplitudes seen in the near-IR, 0.4~mag say, then
the equivalent scatter contributed to the observed PL relation due to
random sampling of the light curve would be 0.40/$\sqrt{12}$ =
$\pm$0.11~mag. If we remove (in quadrature) this random-phase induced
scatter (0.11~mag) from the total measured scatter (0.16~mag) we are
left with $[(0.16)^2 - (0.11)^2]^{1/2}$ = 0.12~mag, and we can simply
draw the following conclusions: Half of the correlation is simply due
to the random sampling of the (highly correlated) light curves, which
at these wavelengths, have equal amplitudes and tightly matched
phases. The other half of the scatter comes from the correlated nature
of the mean-light position of an individual Cepheid within the parent
instability strip: that is, at a given period brighter Cepheids are
brighter at all wavelengths, either due to systematic temperature
differences, radius differences, or both. In either case, the amplitude
of this correlation is expected to be reduced in the mid-IR, but it is
still predicted to be coupled wavelength to wavelength, as is
indeed seen in these datasets.

Figure 3 also illustrates that the (second-order) scatter around the
filter-filter residual plots (for example, $\pm$0.057~mag scatter
about a regression line of slope +0.98$\pm$0.01 in the 3.6 versus
5.8~$\mu$m correlation) is consistent with the range of (individually
plotted) observational errors of $\pm$0.03-0.09~mag given for the
SPITZER data in Table 1.

\vfill\eject
\subsection{Reddening}

From optical studies, estimated reddening values for these LMC
Cepheids range from E(B-V) $=$ 0.00 to 0.20~mag (e.g., Martin, Warren
\& Feast 1979).  The mid-IR extinction curve of the LMC has not been
measured, but measurements of the Galactic extinction curve (Lutz et
al.\ 1996; Indebetouw et al.\ 2005; Flaherty et al.\ 2007;
Roman-Zuniga et al.\ 2007), when scaled from A$_K$ to A$_V$ via Rieke
\& Lebofsky (1985), find A$_{5.8\mu m}$ / A$_K$ $= 0.4$ to $0.5$, and
A$_{5.8\mu m}$ / A$_B$ $\approx 0.04$.


As an example, the most extreme value of E(B-V) given above, 0.20~mag,
yields A$_B = 4.2 \times$ E(B-V) = 0.84~mag.  This converts to
A$_{5.8\mu m}$ =0.03~mag, which is close to the photometric precision
of the current dataset. Most LMC Cepheids have a typical reddening
being perhaps a factor of two smaller (i.e., E(B-V) = 0.10~mag).
Thus, {\it differential reddening effects} around a mean value of
E(B-V) = 0.10~mag would impact the intrinsic calibration of the mid-IR
PL relations for LMC Cepheids at the level of only $\pm$0.01~mag. This
is a huge advantage for measuring precise and accurate extragalactic
distances.

\section{Absolute Calibration}

For the purposes of the present paper, we adopt a true distance to the
Large Magellanic Cloud of (m-M)$_o$ = 18.50~mag and a mean reddening
to the Cepheids of E(B-V) = 0.10~mag (consistent with the values
adopted by the HST Key Project, Freedman {\it et al.}, 2001).
Applying this offset in true distance modulus, subtracting the small
corrections for extinction (0.04 to 0.01~mag), and using simple
least-squares fitting we derive the following provisional absolute
calibrations for the Cepheid Period-Luminosity relations at
mid-infrared wavelengths:

$$M_{3.6} = -3.34 (log(P) - 1.0) ~[\pm 0.02] - 5.87 ~[\pm 0.02] $$

$$M_{4.5} = -3.29 (log(P) - 1.0) ~[\pm 0.03] - 5.92 ~[\pm 0.03] $$

$$M_{5.8} = -3.42 (log(P) - 1.0) ~[\pm 0.03] - 5.83 ~[\pm 0.02] $$

$$M_{8.0} = -3.44 (log(P) - 1.0) ~[\pm 0.03] - 5.89 ~[\pm 0.02] $$

These mid-infrared Period-Luminosity relations and their calibration
hold much promise for improvement. The second-epoch SAGE data is
expected to be released within the next year. Spitzer mid-IR
observations of several nearby Galactic field Cepheids can also
provide an independent {\it absolute} calibration for these PL
relations, based on the geometric parallax distances determined
recently by Benedict {\it et al.} (2007) using the Hubble Space
Telescope Fine-Guidance Sensors. Furthermore, we hope to extend the
observations for this sample to obtain multi-epoch, mid-infrared data
(and time-averaged magnitudes and colors) during the upcoming final
cycle for cold Spitzer observations. In the future, NIRCAM and MIRI on
JWST will be able to provide mid-infrared PL relations for known
Cepheids in nearby galaxies, and a re-calibration of the extragalactic
distance scale.

\section{Conclusions}

We have presented the first absolute calibrations of the Cepheid
Period-Luminosity relations at four mid-infrared wavelengths. These
relations are already good to $\pm$0.02 in the slope and $\pm$0.04~mag
in their relative zero points. At each of these wavelengths the
scatter is such that a single, random-phase observation of a single
Cepheid can provide a distance that is good to $\pm$8\% (statistical
error alone). Given the mid-infrared imaging capabilities of JWST, we are
optimistic that these calibrations can be used to determine
high-precision distances to the most distant galaxies in which
Cepheids have been so far discovered independent of most of the
systematic effects that are currently limiting the accuracy of optical
studies.

\bigskip
\bigskip
\bigskip

We thank the referee for constructive comments, and for suggesting the
inclusion of Figure 4. This research has made use of the NASA/IPAC
Extragalactic Database (NED) which is operated by the Jet Propulsion
Laboratory, California Institute of Technology, under contract with
the National Aeronautics and Space Administration.

{\it Facilities:} \facility{Spitzer}

\vfill\eject
\noindent
\centerline{\bf References \rm}
\vskip 0.1cm
\vskip 0.1cm

\par\noindent
Alves, D. R., 2004, New Astron.Rev., 48,  659

\par\noindent
Benedict, G.~F., {\it et al.} 2007, AJ, 133, 1810

\par\noindent
Bono, G., Caputo, F., Castellani, V., \& Marconi, M. 1999, ApJ, 512, 711

\par\noindent
Cardelli,  J.~A., Clayton, G.~C., \& Mathis, J.~S. 1989, ApJ, 345, 245

\par\noindent
Feast, M.~W., \& Walker, A.~R. 1987, ARAA, 25, 345

\par\noindent
Flaherty, K.~M., Pipher, J.~L., Megeath, S.~T., Winston, E.~M.,
Gutermuth, R.~A.,  Muzerolle, J., Allen, L.~E., \& Fazio, G.~G.,  
2007, ApJ, 663, 1069

\par\noindent
Freedman, W.~L., {\it et al.} 2001, ApJ, 553, 47

\par\noindent
Gibson, B.~K. 2000, Memorie della Societa Astronomica Italiana, 71, 693

\par\noindent
Gieren, W.~P.,  Moffett, T.~J., \&  Barnes, T.~G. III. 1999, ApJ,
512, 553

\par\noindent
Indebetouw,  R., et al., 2005, ApJ, 619, 931

\par\noindent
Lutz,D., et al. 1996, A\&A, 315, L269 

\par\noindent
Madore, B.~F., \& Freedman, W.~L. 1991, PASP, 103, 933

\par\noindent Madore, B.~F., \& Freedman, W.~L. 1998, ``Stellar
Astrophysics for the Local Group'', eds. A. Aparicio, A. Herraro \&
F. Sanchez, Cambridge University Press

\par\noindent
Martin, W.~L., Warren, P.~R., \& Feast, M.~J. 1979, MNRAS, 188 , 139

\par\noindent McGonegal, R., McAlary, C.~W., Madore, B.~F., \&
McLaren, R.~A. 1982, ApJ, 257, L33

\par\noindent
Meixner, M., {\it et al.} 2006a, AJ, 132, 2268

\par\noindent
Meixner, M., {\it et al.} 2006b, ``The SAGE Data Description:
Delivery 1''
\par
http://sage.stsci.edu/SAGE$\_$SSCdatadocument$\_$v5.pdf

\par\noindent
Persson, S.~E., {\it et al.} 2004, AJ, 128, 2239

\par\noindent
Rieke, G.~H., \& Lebofsky, M.~J. 1985, ApJ, 288, 618

\par\noindent
Riess, A.~G., {\it et al.}, 2005, ApJ, 627, 579

\par\noindent
Rom{\'a}n-Z{\'u}{\~n}iga, C.~G., Lada, C.~J., Muench, A., \& Alves,   
J.~F.\
2007, ApJ, 664, 357

\par\noindent Sandage, A.~R., Tammann, G.~A., Saha, A., Reindl, B.,
Macchetto, F.~D., \& Panagia, N. 2006, ApJ, 653, 843

\par\noindent
Schaefer, B.E. 2007, AJ, (in press) =  arXiv:0709.4531

\par\noindent
Udalski, A., Szymanski, M., Kubiak, M., Pietrzynski, G., Soszynski, I,
Wozniak, P., \& Zebrun, K.  1999, Acta Astronomica, 49, 201

\par\noindent
Wisnewski, W.~Z., \& Johnson, H.~L. 1968, Comm. Lunar Planet. Lab,
No. 112
\par\noindent

\vskip 0.75cm

\clearpage



\begin{figure}
\epsscale{.80}
\plotone{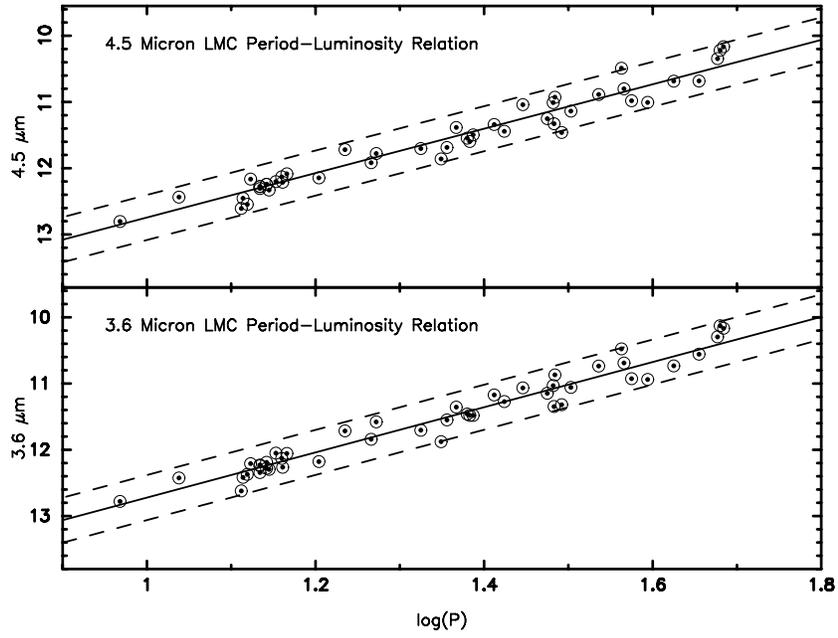}
\caption{IRAC 3.6 and 4.5 $\mu$m Period-Luminosity relations
for LMC Cepheids in the log(P) range from 0.8 to 1.8. The broken lines
represent $\pm$2-$\sigma$ ($\pm$0.32~mag) bounds on the instability
strip.}
\end{figure}

\begin{figure}
\epsscale{.80}
\plotone{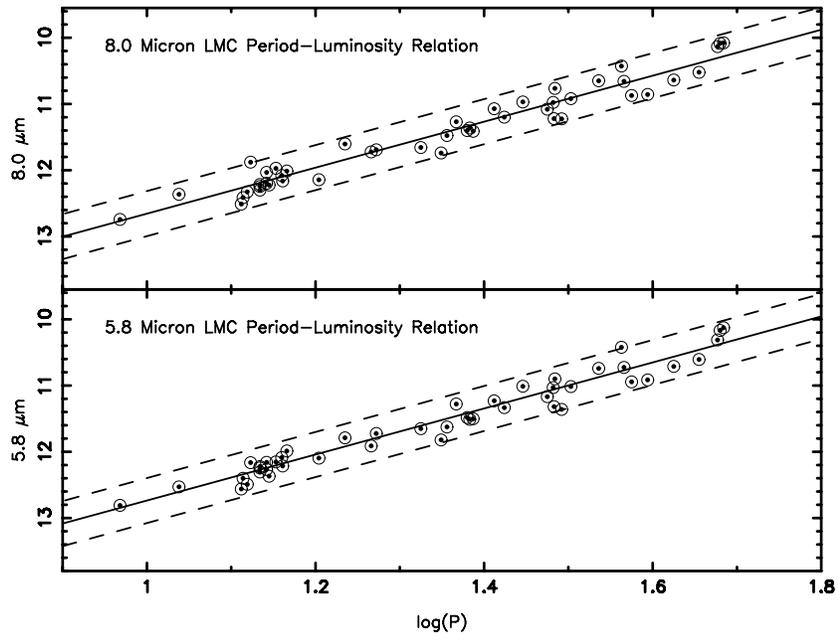}
\caption{IRAC 5.8 and 8.0 $\mu$m Period-Luminosity relations
for LMC Cepheids in the log(P) range from 0.8 to 1.8.}
\end{figure}

\clearpage

\begin{figure}
\epsscale{.80}
\plotone{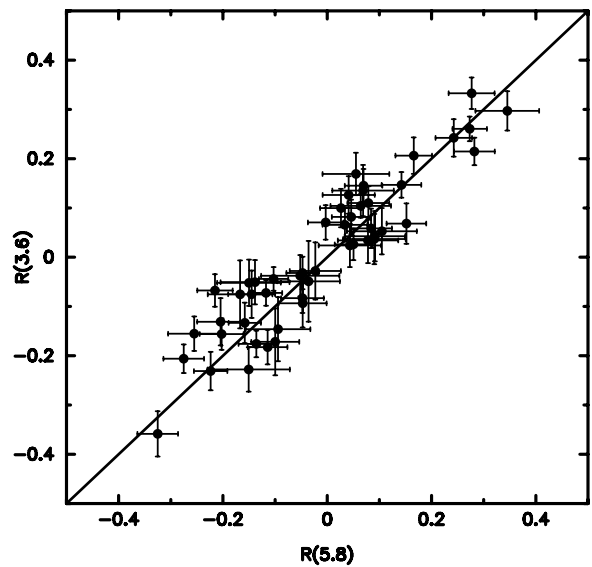}
\caption{Residuals in the PL fit at 3.6~$\mu$m
versus residuals in the PL fit at 5.8~$\mu$m, illustrating both the
high degree of correlation between the residuals, accounting for most
of the variance, and the small residual scatter in the correlation.
The correlation is driven by the random-phase nature of the sampling
of the Cepheid light curves in addition to the correlated
nature of the mean properties of Cepheids within and across the
instability strip. The residual noise about the mean correlation is
entirely consistent with the quoted mean photometric errors in the
individual data points (i.e., 0.05~mag).}
\end{figure}

\begin{figure}
\epsscale{.80}
\plotone{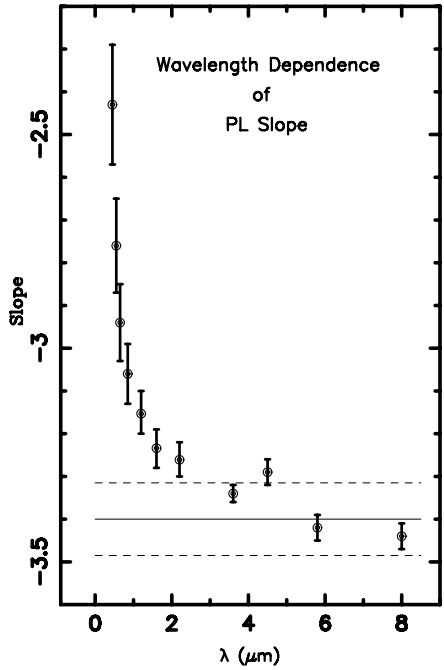}
\caption{The steepening of the slope of the Cepheid Period-Luminosity
relation as a function of increasing wavelength, from the optical
(BVRI) through the near infrared (JHK) and out to 8 microns in the
IRAC mid-infrared.  An asymptotic value (predicted from the
Period-Radius relation) and its one-sigma uncertainties are shown by
the horizontal lines at -3.40 $\pm$ 0.085.}
\end{figure}

\clearpage`

\clearpage

\begin{deluxetable}{lccccc}
\tablecolumns{6}
\tablewidth{4.5truein}
\tablehead{
\colhead{Cepheid}  & \colhead{~~~~~log(P)~~~~~}  & \colhead{3.6$\mu 
$m}    & \colhead{4.5$\mu$m}  & \colhead{5.8$\mu$m } & \colhead{8.0$ 
\mu$m }
\\ \colhead{}    & \colhead{(days)}    & \colhead{(mag)} & \colhead 
{(mag)}   & \colhead{(mag)} &\colhead{(mag)} \cr
}

\startdata
HV 872   & 1.475 & 11.15 & 11.25 & 11.17 & 11.08 \\
       & & 0.04 & 0.05 & 0.04 & 0.04  \\
HV 873   & 1.536 & 10.74 & 10.89 & 10.74 & 10.65 \\
       & & 0.04 & 0.05 & 0.03 & 0.04  \\
HV 875   & 1.482 & 11.03 & 11.01 & 11.03 & 10.98 \\
       & & 0.03 & 0.04 & 0.03 & 0.03 \\
HV 876   & 1.356 & 11.55 & 11.69 & 11.63 & 11.48 \\
       & & 0.04 & 0.05 & 0.05 & 0.05  \\
HV 877   & 1.655 & 10.56 & 10.68 & 10.61 & 10.52 \\
       & & 0.03 & 0.05 & 0.04 & 0.04 \\
HV 878   & 1.367 & 11.36 & 11.38 & 11.28 & 11.27 \\
       & & 0.04 & 0.06 & 0.04 & 0.03 \\
HV 879   & 1.566 & 10.69 & 10.80 & 10.73 & 10.66 \\
       & & 0.02 & 0.04 & 0.03 & 0.03 \\
HV 882   & 1.503 & 11.06 & 11.14 & 11.02 & 10.92 \\
       & & 0.03 & 0.05 & 0.04 & 0.03 \\
HV 886   & 1.380 & 11.46 & 11.55 & 11.48 & 11.40 \\
       & & 0.03 & 0.05 & 0.04 & 0.04 \\
HV 887   & 1.161 & 12.26 & 12.21 & 12.21 & 12.16 \\
       & & 0.05 & 0.08 & 0.06 & 0.04 \\
HV 889   & 1.412 & 11.17 & 11.34 & 11.23 & 11.07 \\
       & & 0.05 & 0.09 & 0.05 & 0.04 \\
HV 891   & 1.235 & 11.72 & 11.72 & 11.79 & 11.60 \\
       & & 0.05 & 0.06 & 0.05 & 0.04 \\
HV 892   & 1.204 & 12.18 & 12.15 & 12.10 & 12.14 \\
       & & 0.04 & 0.05 & 0.04 & 0.05 \\
HV 893   & 1.325 & 11.70 & 11.70 & 11.65 & 11.66 \\
       & & 0.05 & 0.05 & 0.06 & 0.04 \\
HV 899   & 1.492 & 11.32 & 11.46 & 11.36 & 11.23 \\
       & & 0.03 & 0.06 & 0.04 & 0.04 \\
HV 900   & 1.677 & 10.30 & 10.35 & 10.31 & 10.14 \\
       & & 0.03 & 0.04 & 0.03 & 0.03 \\
HV 901   & 1.266 & 11.84 & 11.92 & 11.91 & 11.72 \\
       & & 0.04 & 0.06 & 0.04 & 0.05 \\
HV 904   & 1.483 & 11.35 & 11.33 & 11.32 & 11.22 \\
       & & 0.03 & 0.03 & 0.03 & 0.03 \\
HV 909   & 1.575 & 10.93 & 10.98 & 10.94 & 10.87 \\
       & & 0.04 & 0.06 & 0.04 & 0.04 \\
HV 914   & 0.838 & 13.02 & 13.06 & 13.14 & 12.88 \\
       & & 0.04 & 0.08 & 0.05 & 0.07 \\
HV 932   & 1.123 & 12.21 & 12.16 & 12.16 & 11.88 \\
       & & 0.06 & 0.07 & 0.06 & 0.10 \\
HV 953   & 1.680 & 10.13 & 10.22 & 10.17 & 10.09 \\
       & & 0.03 & 0.05 & 0.04 & 0.03 \\
HV 971   & 0.968 & 12.78 & 12.80 & 12.81 & 12.74 \\
       & & 0.04 & 0.08 & 0.08 & 0.07 \\
HV 997   & 1.119 & 12.37 & 12.55 & 12.49 & 12.33 \\
       & & 0.04 & 0.06 & 0.06 & 0.05 \\
HV 1002  & 1.484 & 10.87 & 10.93 & 10.90 & 10.76 \\
       & & 0.03 & 0.04 & 0.04 & 0.03 \\
HV 1003  & 1.387 & 11.48 & 11.50 & 11.50 & 11.41 \\
       & & 0.03 & 0.05 & 0.04 & 0.04 \\
HV 1005  & 1.272 & 11.58 & 11.78 & 11.72 & 11.69 \\
       & & 0.03 & 0.04 & 0.03 & 0.03 \\
HV 1006  & 1.153 & 12.05 & 12.20 & 12.15 & 11.97 \\
       & & 0.05 & 0.06 & 0.06 & 0.04 \\
HV 1013  & 1.383 & 11.48 & 11.60 & 11.51 & 11.36 \\
       & & 0.02 & 0.04 & 0.06 & 0.03 \\
HV 1019  & 1.134 & 12.24 & 12.28 & 12.24 & 12.30 \\
       & & 0.06 & 0.06 & 0.05 & 0.07 \\
HV 1023  & 1.424 & 11.27 & 11.44 & 11.33 & 11.20 \\
       & & 0.04 & 0.04 & 0.03 & 0.03 \\
HV 2244  & 1.145 & 12.30 & 12.33 & 12.37 & 12.22 \\
       & & 0.05 & 0.06 & 0.06 & 0.05 \\
HV 2251  & 1.446 & 11.06 & 11.04 & 11.01 & 10.97 \\
       & & 0.03 & 0.06 & 0.03 & 0.03 \\
HV 2257  & 1.594 & 10.94 & 11.01 & 10.91 & 10.86 \\
       & & 0.04 & 0.06 & 0.04 & 0.04 \\
HV 2260  & 1.114 & 12.42 & 12.45 & 12.40 & 12.41 \\
       & & 0.04 & 0.05 & 0.04 & 0.04 \\
HV 2270  & 1.134 & 12.34 & 12.31 & 12.31 & 12.22 \\
       & & 0.05 & 0.05 & 0.06 & 0.05 \\
HV 2279  & 0.839 & 13.12 & 13.08 & 13.07 & 13.01 \\
       & & 0.05 & 0.06 & 0.08 & 0.09 \\
HV 2282  & 1.166 & 12.06 & 12.08 & 11.99 & 12.01 \\
       & & 0.07 & 0.06 & 0.05 & 0.05 \\
HV 2291  & 1.349 & 11.88 & 11.86 & 11.82 & 11.74 \\
       & & 0.04 & 0.07 & 0.06 & 0.04 \\
HV 2294  & 1.563 & 10.48 & 10.49 & 10.42 & 10.43 \\
       & & 0.05 & 0.05 & 0.04 & 0.04 \\
HV 2324  & 1.160 & 12.13 & 12.13 & 12.09 & 12.08 \\
       & & 0.05 & 0.05 & 0.05 & 0.04 \\
HV 2337  & 0.837 & 13.14 & 13.19 & 13.25 & 13.15 \\
       & & 0.05 & 0.10 & 0.07 & 0.08 \\
HV 2338  & 1.625 & 10.73 & 10.69 & 10.71 & 10.64 \\
       & & 0.03 & 0.05 & 0.04 & 0.03 \\
HV 2339  & 1.142 & 12.28 & 12.24 & 12.27 & 12.20 \\
       & & 0.04 & 0.05 & 0.06 & 0.05 \\
HV 2352  & 1.134 & 12.23 & 12.27 & 12.22 & 12.24 \\
       & & 0.08 & 0.05 & 0.06 & 0.07 \\
HV 2369  & 1.684 & 10.17 & 10.17 & 10.13 & 10.08 \\
       & & 0.04 & 0.05 & 0.03 & 0.04 \\
HV 2405  & 0.840 & 13.37 & 13.33 & 13.35 & 13.32 \\
       & & 0.05 & 0.05 & 0.07 & 0.08 \\
HV 2432  & 1.038 & 12.43 & 12.44 & 12.53 & 12.36 \\
       & & 0.07 & 0.05 & 0.06 & 0.05 \\
HV 2527  & 1.112 & 12.62 & 12.61 & 12.56 & 12.51 \\
       & & 0.03 & 0.06 & 0.04 & 0.06 \\
HV 2538  & 1.142 & 12.19 & 12.24 & 12.16 & 12.03 \\
       & & 0.03 & 0.03 & 0.04 & 0.03 \\
HV 2549  & 1.209 & 11.71 & 11.76 & 11.74 & 11.67 \\
       & & 0.04 & 0.06 & 0.05 & 0.04 \\
HV 2579  & 1.128 & 12.08 & 12.12 & 12.08 & 12.02 \\
       & & 0.04 & 0.05 & 0.05 & 0.06 \\
HV 2580  & 1.228 & 11.69 & 11.82 & 11.76 & 11.66 \\
       & & 0.03 & 0.03 & 0.03 & 0.03 \\
HV 2733  & 0.941 & 12.86 & 12.82 & 12.85 & 12.76 \\
       & & 0.05 & 0.05 & 0.05 & 0.07 \\
HV 2749  & 1.364 & 11.69 & 11.77 & 11.69 & 11.49 \\
       & & 0.04 & 0.06 & 0.05 & 0.05 \\
HV 2793  & 1.283 & 11.50 & 11.62 & 11.56 & 11.47 \\
       & & 0.03 & 0.06 & 0.05 & 0.04 \\
HV 2836  & 1.244 & 11.93 & 12.03 & 11.91 & 12.01 \\
       & & 0.03 & 0.05 & 0.05 & 0.04 \\
HV 2854  & 0.936 & 12.91 & 12.88 & 12.94 & 12.83 \\
       & & 0.03 & 0.04 & 0.04 & 0.05 \\
HV 5655  & 1.153 & 12.14 & 12.18 & 12.11 & 12.06 \\
       & & 0.05 & 0.06 & 0.05 & 0.05 \\
HV 6065  & 0.835 & 13.17 & 13.26 & 13.16 & 13.28 \\
       & & 0.04 & 0.06 & 0.07 & 0.08 \\
HV 6098  & 1.384 & 11.14 & 11.13 & 11.14 & 11.10 \\
       & & 0.03 & 0.04 & 0.03 & 0.02 \\
HV 8036  & 1.453 & 11.19 & 11.31 & 11.22 & 11.24 \\
       & & 0.04 & 0.06 & 0.04 & 0.04 \\
HV 12471 & 1.200 & 12.00 & 12.13 & 12.03 & 11.96 \\
       & & 0.04 & 0.05 & 0.05 & 0.05 \\
HV 12505 & 1.158 & 12.39 & 12.45 & 12.53 & 12.41 \\
       & & 0.04 & 0.05 & 0.08 & 0.05 \\
HV 12656 & 1.127 & 12.31 & 12.30 & 12.29 & 12.25 \\
       & & 0.05 & 0.05 & 0.05 & 0.05 \\
HV 12700 & 0.911 & 12.88 & 12.90 & 12.99 & 12.85 \\
       & & 0.04 & 0.06 & 0.07 & 0.06 \\
HV 12724 & 1.138 & 12.29 & 12.36 & 12.34 & 12.27 \\
       & & 0.05 & 0.05 & 0.06 & 0.05 \\
HV 12815 & 1.416 & 11.03 & 11.12 & 11.06 & 10.96 \\
       & & 0.04 & 0.05 & 0.04 & 0.03 \\
HV 12816 & 0.973 & 12.87 & 12.84 & 12.90 & 12.91 \\
       & & 0.05 & 0.05 & 0.07 & 0.06 \\
HV 13048 & 0.836 & 13.06 & 13.09 & 13.07 & 13.00 \\
       & & 0.03 & 0.05 & 0.05 & 0.06 \\
\enddata
\end{deluxetable}

\vfill\eject
\noindent

\end{document}